\documentclass[aps,print,showpacs,twocolumn]{revtex4}
\usepackage{graphicx}
\usepackage{amsfonts}
\usepackage{amsmath}
\usepackage{amssymb}

\begin{document}

\title{Geometric phase in the Kitaev honeycomb model and scaling behavior at
critical points}
\author{Jinling Lian}
\affiliation{Institute of Theoretical Physics, Shanxi University, Taiyuan 030006, China}
\author{J.-Q. Liang}
\thanks{jqliang@sxu.edu.cn}
\affiliation{Institute of Theoretical Physics, Shanxi University, Taiyuan 030006, China}
\author{Gang Chen}
\thanks{chengang971@163.com}
\affiliation{State Key Laboratory of Quantum Optics and Quantum Optics Devices, Shanxi
University, Taiyuan 030006, China}
\affiliation{Department of Physics, Shaoxing University, Shaoxing 312000, China}

\begin{abstract}
In this paper a geometric phase of the Kitaev honeycomb model is derived and
proposed to characterize the topological quantum phase transition. The
simultaneous rotation of two spins is crucial to generate the geometric
phase for the multi-spin in a unit-cell unlike the one-spin case. It is
found that the ground-state geometric phase, which is non-analytic at the
critical points, possesses zigzagging behavior in the gapless $B$ phase of
non-Abelian anyon excitations, but is a smooth function in the gapped $A$
phase. Furthermore, the finite-size scaling behavior of the non-analytic
geometric phase along with its first- and second-order partial derivatives
in the vicinity of critical points is shown to exhibit the universality. The
divergent second-order derivative of geometric phase in the thermodynamic
limit indicates the typical second-order phase transition and thus the
topological quantum phase transition can be well described in terms of the
geometric-phase.
\end{abstract}

\volumeyear{year}
\volumenumber{number}
\issuenumber{number}
\eid{identifier}
\received[Received text]{date}
\revised[Revised text]{date}
\accepted[Accepted text]{date}
\published[Published text]{date}
\maketitle


\section{Introduction}

Berry in his pioneer work raised a fundamentally important concept known as
geometric phase (GP) in addition to the usual dynamic phase accumulated on
the wave function of a quantum system, provided that the Hamiltonian varies
with multi-parameters cyclically and adiabatically \cite{Berry}. At the
present time the GP with extensive generalization along many directions has
wide applications in various branches of physics \cite{Berry2,Shapere,Bohm}.

Recently, the close relation between GP and quantum phase transition (QPT)
has been gradually revealed \cite{Carollo,Zhu,Hamma} and increasing interest
has been drawn to the role of GP in detecting QPT for various many-body
systems \cite{Zhu08,Chen,interests}, which, as a matter of fact, is also a
new research field in condensed matter physics \cite{Sachdev,Sondhi}. QPT
usually describes an abrupt change in the ground state of a many-body system
induced by quantum fluctuations. The phase transition between ordered and
disordered phases is accompanied by symmetry breaking, which can also be
characterized by Landau-type order parameters.

On the other hand, a new type of QPT called topological quantum phase
transitions (TQPT) has attracted much attention. The first non-trivial
example is the fractional quantum Hall effect \cite{Tsui,Laughlin}. In the
last decade, several exactly soluble spin-models with the TQPT, such as the
toric-code model \cite{Kitaev03}, the Wen-plaquette model \cite%
{Wen-plaquette,Wen08} and the Kitaev model on a honeycomb lattice \cite%
{Kitaev}, were found. In contrast to the conventional QPT governed by local
order parameters \cite{Sachdev}, the TQPT can be characterized only by the
topological order \cite{WenBook}. As good examples to illustrate the
underlying physics, different methods are developed to describe the TQPT in
the Kitaev honeycomb model \cite{T. Xiang07,T. Xiang08,H. D. Chen07,H. D.
Chen08,S. Yang,Gu}. In Ref. \cite{T. Xiang07}, Feng \textit{et.al.} obtained
the local order parameters of Landau type to characterize the phase
transition by introducing Jordan-Wigner and spin-duality transformations
into the Majorana representation of the honeycomb model. Gu \textit{et.al.}
showed an exciting result of the ground-state fidelity susceptibility \cite%
{S. Yang,Gu}, which can be used to identify the TQPT from the gapped $A$
phase with Abelian anyon excitations to gapless $B$ phase with non-Abelian
anyon excitations.

Quite recently, Zhu \cite{Zhu} showed that the ground-state GP in the $XY$
model is non-analytic with a diverged derivative with respect to the field
strength at the critical value of magnetic field. Thereupon, the relation
between the GP and the QPT is established. Nevertheless, much attention has
been paid to the QPT, while effort devoted to the relation between the GP
and the TQPT is very little. The present paper is devoted to exploiting the
GP of the Kitaev honeycomb model as an essential tool to establish a
relation between the GP and the TQPT and reveal the novel quantum
criticality. Unlike the GP in the usual lattice-spin model for the QPT,
which is generated by the single-spin rotation of each lattice-site, the
simultaneous rotation of linked two spins in one unit-cell seems crucial to
describe the TQPT in the honeycomb model. The non-analyticity of GP at the
critical points with a divergent second-order derivative with respect to the
coupling parameters shows that the TQPT is the second-order transition and
can be well described by the GP.

In Sec. II, the ground state wave function and energy spectrum of the Kitaev
honeycomb model are presented. After introducing a correlated rotation of
two $z$-link spins in each unit-cell, the ground-state GP and its
derivatives are obtained explicitly in Sec. III. Sec. IV is devoted to
investigating the scaling behavior of the GP. A brief summary and discussion
are given in Sec. V.

\section{The Kitaev honeycomb model and spectrum}

\begin{figure}[h]
\centering \vspace{0cm} \hspace{0cm} \scalebox{0.8}{%
\includegraphics{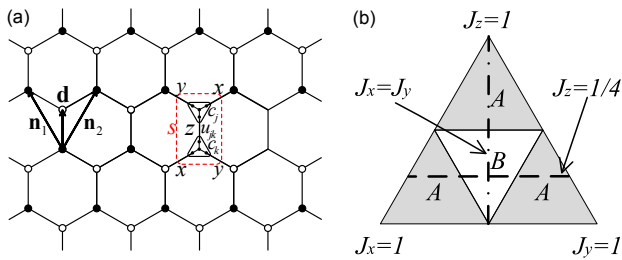}}
\caption{(Color online) (a) Kitaev honeycomb model, in which one spin
interacts with three nearest-neighbor spins through three types of bonds,
depending on their direction. A unit-cell with $x$, $y$ and $z$ links and
graphic representation of Hamiltonian (\protect\ref{HM}) with Majorana
operators are marked by the red dotted line. (b) Phase diagram of the model,
where $A$ phase is gapped and $B$ phase is gapless.}
\label{fig1}
\end{figure}

The Kitaev honeycomb model shown in Fig. \ref{fig1}(a) is firstly introduced
to illustrate the topologically fault-tolerant quantum-information
processing \cite{Kitaev03,Kitaev,Nayak}. In this model, each spin located at
vertices of the lattice interacts with three nearest-neighbor spins through
three types of bonds, depending on their directions. By using the Pauli
operators $\sigma ^{a}$ $(a=x,y,z)$, the corresponding Hamiltonian is
written as%
\begin{equation}
H=-J_{x}\!\!\!\sum_{x\mbox{-links}}\!\!\!\sigma _{j}^{x}\sigma
_{k}^{x}-J_{y}\!\!\!\sum_{y\mbox{-links}}\!\!\!\sigma _{j}^{y}\sigma
_{k}^{y}-J_{z}\!\!\!\sum_{z\mbox{-links}}\!\!\!\sigma _{j}^{z}\sigma
_{k}^{z},  \label{Kitaev model}
\end{equation}%
where $j$, $k$ denote the two ends of the corresponding bond, and $J_{a}$
are coupling parameters. After introducing a special notation $K_{jk}=\sigma
_{j}^{a}\sigma _{k}^{a}$, where the indexes $a$ depend on the types of links
between sites $j$ and $k$ (so we also write it into $a_{jk}$ in the
following text for perspicuousness), Hamiltonian (\ref{Kitaev model}) can be
rewritten into a compact form
\begin{equation}
H=-\frac{1}{2}\sum_{\langle j,k\rangle }J_{a_{jk}}K_{jk}.
\label{Kitaev modelK}
\end{equation}

It has been known that the Kitaev honeycomb model can be solved exactly by
introducing Majorana fermion operators, which are defined as \cite{Kitaev,S.
Yang}
\begin{equation}
\sigma ^{x}=ib^{x}c,\quad \sigma ^{y}=ib^{y}c,\quad \sigma ^{z}=ib^{z}c.
\label{MajoranaFermion}
\end{equation}%
Generally, a set of Majorana operators $M=\left\{
b^{x},b^{y},b^{z},c\right\} $ can be employed to describe a spin by two
fermionic modes. They are Hermitian and obey the relations $m^{2}=1$ and $%
mm^{\prime }=-m^{\prime }m$ for $m,m^{\prime }\in M$ and $m\neq m^{\prime }$%
. Moreover, in the Hilbert space with a spin described by two fermionic
modes, the relation $b^{x}b^{y}b^{z}c\left\vert \Psi \right\rangle
=\left\vert \Psi \right\rangle $\ must be satisfied to ensure the obeying of
the same algebraic relations as $\sigma^{x}$, $\sigma ^{y}$, and $\sigma
^{z} $ \cite{Kitaev}.

Drawing on the operators (\ref{MajoranaFermion}) to the Kitaev honeycomb
model, Hamiltonian (\ref{Kitaev modelK}) is given by%
\begin{equation}
H=\frac{i}{2}\sum_{\langle j,k\rangle}\hat{u}_{jk}J_{a_{jk}}c_{j}c_{k},
\label{HM}
\end{equation}
where $\hat{u}_{jk}=ib_{j}^{a_{jk}}b_{k}^{a_{jk}}$. Fig. \ref{fig1}(a) also
shows the structure of Hamiltonian (\ref{HM}), from which it can be seen
that $\hat{u}_{jk}=-\hat{u}_{kj}$. Since these operators $\hat{u}_{jk}$
commute with the Hamiltonian (\ref{HM}) and with each other, the Hilbert
space splits into two common eigenspaces of $\hat{u}_{jk}$ with eigenvalues $%
u_{jk}=\pm1$. Thus, Hamiltonian (\ref{HM}) is reduced to a quadratic
Majorana fermionic Hamiltonian
\begin{equation}
H=\frac{i}{2}\sum_{\langle j,k\rangle}u_{jk}J_{a_{jk}}c_{j}c_{k}.  \label{HU}
\end{equation}

With a Fourier transformation
\begin{equation}
c_{s,\lambda}=\frac{1}{\sqrt{2L^{2}}}\sum_{\mathbf{q}}e^{i\mathbf{q}\cdot%
\mathbf{r}_{s}}a_{\mathbf{q},\lambda},  \label{Fourier}
\end{equation}
where $s$ denotes a unit cell shown in Fig. \ref{fig1}(a), $\lambda $ refers
to a position inside the cell, $r_{s}$ represents the coordinate of the unit
cell, and $\mathbf{q}$ are momenta of the system with finite system-size $%
2L^{2}$, and a Bogoliubov transformation
\begin{equation}
\left\{
\begin{array}{c}
C_{\mathbf{q},1}^{\dag}=\frac{1}{\sqrt{2}}a_{-\mathbf{q},1}-\frac{1}{\sqrt{2}%
}A_{\mathbf{q}}^{\ast}a_{-\mathbf{q},2}, \\
C_{\mathbf{q},2}^{\dag}=\frac{1}{\sqrt{2}}A_{\mathbf{q}}a_{-\mathbf{q},1}+%
\frac{1}{\sqrt{2}}a_{-\mathbf{q},2},%
\end{array}
\right.
\end{equation}
where $A_{\mathbf{q}}=\sqrt{\epsilon_{\mathbf{q}}^{2}+\Delta_{\mathbf{q}}^{2}%
}/(\Delta_{\mathbf{q}}+i\epsilon_{\mathbf{q}})$, Hamiltonian (\ref{HU}) is
transformed into%
\begin{equation}
H=\sum_{\mathbf{q}}\sqrt{\epsilon_{\mathbf{q}}^{2}+\Delta_{\mathbf{q}}^{2}}%
\left( C_{\mathbf{q},1}^{\dagger}C_{\mathbf{q},1}-C_{\mathbf{q},2}^{\dagger
}C_{\mathbf{q},2}\right)  \label{HC}
\end{equation}
with $\epsilon_{\mathbf{q}}=J_{x}\cos q_{x}+J_{y}\cos q_{y}+J_{z}$, and $%
\Delta_{\mathbf{q}}=J_{x}\sin q_{x}+J_{y}\sin q_{y}$. In Hamiltonian (\ref%
{HC}), the momenta take the values \cite{S. Yang}%
\begin{equation}
q_{x\left( y\right) }=\frac{2n\pi}{L},n=-\frac{L-1}{2},\cdots,\frac{L-1}{2},
\label{Values of q}
\end{equation}
when the system size is chosen as $N=2L^{2}$ with $L$ being an odd integer.
Thus, the ground and the first-excited states are obtained by

\begin{align}
\left\vert \Psi_{0}\right\rangle & \!\!=\prod_{\mathbf{q}}C_{q,2}^{\dagger
}\left\vert 0\right\rangle =\prod_{\mathbf{q}}\frac{1}{\sqrt{2}}\left( A_{%
\mathbf{q}}a_{-\mathbf{q},1}+a_{-\mathbf{q},2}\right) \left\vert
0\right\rangle ,  \label{wavefunction} \\
\left\vert \Psi_{1}\right\rangle & \!\!=\prod_{\mathbf{q}}C_{q,1}^{\dagger
}\left\vert 0\right\rangle =\prod_{\mathbf{q}}\frac{1}{\sqrt{2}}\left( a_{-%
\mathbf{q},1}-A_{\mathbf{q}}^{\ast}a_{-\mathbf{q},2}\right) \! \left\vert
0\right\rangle ,
\end{align}
with the energy eigenvalues%
\begin{equation}
E_{0,1}=\pm\sum_{\mathbf{q}}\sqrt{\epsilon_{\mathbf{q}}^{2}+\Delta _{\mathbf{%
q}}^{2}},  \label{EP}
\end{equation}

\begin{figure}[t]
\centering \vspace{0cm} \hspace{0cm} \scalebox{0.88}{%
\includegraphics{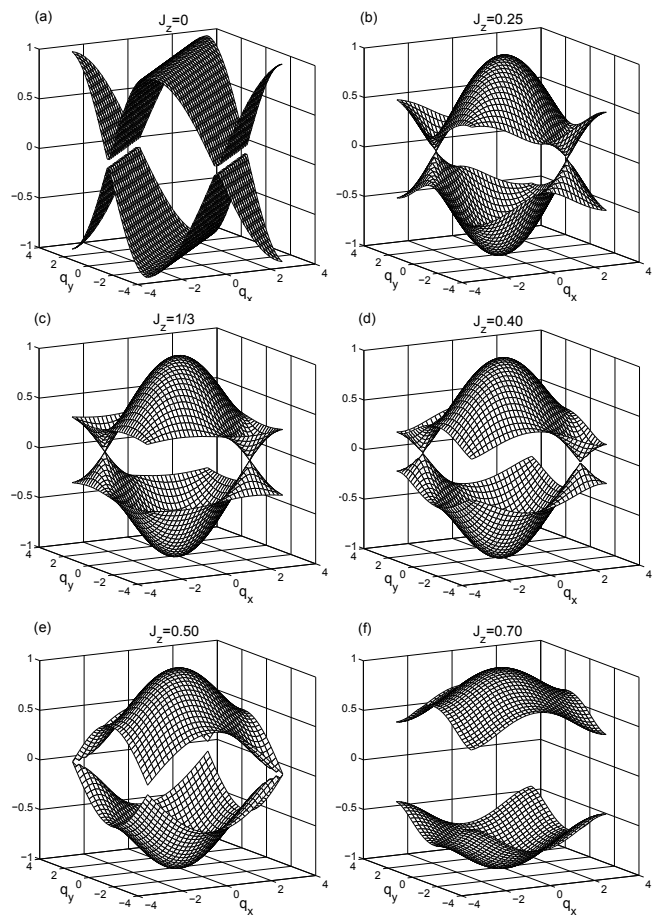}}.
\caption{Energy spectrum for the parameters $J_{x}=J_{y}$ (a) $J_{z}=0$, the
spectrum indeed is degenerate in a larger system size; (b), (c), and (d) in $%
B$ phase, for $J_{z}<1/3$, $J_{z}=1/3$, and $1/3<J_{z}<1/2$, respectively;
(e) at the critical point of $J_{z}=1/2$ and (f) in $A$ phase. It is clearly
that there is level-crossing in (b), (c) and (d).}
\label{fig2}
\end{figure}

It has been shown that the Kitaev honeycomb model (\ref{Kitaev model}) has a
rich phase diagram including a gapped phase with Abelian anyonic excitations
(called $A$ phase) and a gapless phase with non-Abelian anyonic excitations (%
$B$ phase) \cite{Kitaev}. In Fig. \ref{fig1}(b), the two phases $A $ and $B$
are separated by three transition lines, \textit{i.e.}, $J_{x}=1/2$, $%
J_{y}=1/2$, and $J_{z}=1/2$, which form a small triangle surrounding the $B$
phase. Here, we only plot the energy spectrum (\ref{EP}) as a function of $%
J_{z}$ for $J_{x}=J_{y}$ (the vertical dot-and-dash line in Fig. \ref{fig1}%
(b)) in Fig. \ref{fig2}. It can be seen from Fig. \ref{fig2} that the
energy-level degeneracy arises or lifts at certain points, which can be
regarded as the possible critical points of QPT \cite{Zhu,Sachdev}. In Fig. %
\ref{fig2}(b, c, d), the degenerate points occur in the $B$ phase, but
disappear in the $A$ phase as shown in Fig. \ref{fig2}(f). Moreover, the
energy spectrum may have asymptotic degeneracy at the phase diagram edge
seen from Fig. \ref{fig2}(a) when the size of system tends to infinity. The
non-analyticity points of ground state in $B$ phase are actual
level-crossing points.

\section{Geometric phase}

In general spin-chain systems characterized by Landau-type order parameters,
the GPs can be generated by adiabatic and cyclic evolutions with a rotation
of each spin coupled with its neighbors in the certain parameter spaces \cite%
{Zhu}. However, in the Kitaev honeycomb model, the two spins $%
\sigma_{s,\lambda }$ and $\sigma _{s,\bar{\lambda}}$ in the unit cell $s$ is
coupled with each other only by a $z$-link, as Fig. \ref{fig1}(a) shows. As
a result, a simple rotation of a single-spin with the generator $\sigma
_{s,\lambda }^{z}$ can only "propagate" along the horizontal spin-chain
through the $x$- and $y$-link couplings but not along the vertical direction
since the rotation-generator of a single-spin $\sigma _{s,\lambda }^{z}$
commutes with the $z$-link spin-operators in the Hamiltonian between the
horizontal spin-chains. On other words, if using the above single-spin
rotation, the Kitaev honeycomb model would be equivalent to a system of
independent multi-single-spin-chains. In order to obtain the GP of the
Kitaev honeycomb model, it should be introduced a correlated rotation of two
$z$-link spins in each unit-cell,%
\begin{equation}
U\left( \phi \right) =\exp \left[ i\phi R\right] ,  \label{Rotation}
\end{equation}%
where%
\begin{equation}
R=\sum_{s}\sum_{\lambda }\sigma _{s,\lambda }^{z}\sigma _{s,\bar{\lambda}%
}^{z}  \label{R expression}
\end{equation}%
is the correlated rotation-generator, and $\phi $ is the "co-rotation
angle". Indeed, when the rotation operator is applied on the spin of $\left(
s,\lambda \right) $-th set, the spin operator $\sigma _{s,\bar{\lambda}}^{z}$
acts as a c-number because the spin operators on different sets commute each
other. As a matter of fact, the correlated rotation-generator $R$ can
generate a "co-rotation" of the $z$-link spins in each unit cell.
Furthermore, under the gauge transformation (\ref{Rotation}), the rotations
of each spins in the 2D honeycomb Kitaev model become correlative and
global. Hence it is possible to reveal the relation between the GP and the
nonlocal TQPT by means of the correlated rotation. Then, the ground state
wave function becomes
\begin{equation}
\left\vert \Psi _{0}^{\prime }(\phi )\right\rangle =U\left( \phi \right)
\left\vert \Psi _{0}\right\rangle ,
\end{equation}%
and correspondingly, the GP is given by \cite{Liang}%
\begin{align}
\gamma & =-i{\int\nolimits_{0}^{2\pi }}\left\langle \Psi _{0}\right\vert
U^{^{\dag }}\frac{d}{d\phi }U\left\vert \Psi _{0}\right\rangle d\phi  \notag
\\
& =-2\pi \left\langle \Psi _{0}\right\vert R\left\vert \Psi
_{0}\right\rangle .  \label{BPOrigin}
\end{align}%
It can be seen that the GP for the Kitaev honeycomb model is proportional to
the expectation value of the correlated rotation $R$. Different from the
local spin-correlation $\sigma _{s,\lambda }^{z}\sigma _{s,\bar{\lambda}%
}^{z} $, the sum over whole lattice-sites in $R$ leads to a global property,
which is crucial for the topological QPT.

In terms of the Majorana fermion operators (\ref{MajoranaFermion}), the
rotation operator $R$ is given, after using the Fourier transformation (\ref%
{Fourier}), by $R=\sum_{s,\lambda}\left( ib_{s,\lambda}^{z}c_{s,\lambda
}\right) \left( ib_{s,\bar{\lambda}}^{z}c_{s,\bar{\lambda}}\right)
=\sum_{s}\left( -iu_{s,1;s,2}^{z}c_{s,1}c_{s,2}\right) +\sum_{s}\left(
-iu_{s,2;s,1}^{z}c_{s,2}c_{s,1}\right) $. Under the restriction of vortex
free subspace, $u_{s,1;s,2}^{z}$ and $u_{s,2;s,1}^{z}$ are the good quantum
numbers, that is to say, $u_{s,1;s,2}^{z}=1$, and $u_{s,2;s,1}^{z}=-1$.
Thus, the rotation operator $R$ is finally obtained by

\begin{equation}
R=\frac{-i}{\sqrt{2L^{2}}}\sum_{\mathbf{q}}\left( a_{-\mathbf{q},1}a_{%
\mathbf{q},2}-a_{-\mathbf{q},2}a_{\mathbf{q},1}\right) .  \label{R}
\end{equation}
By means of Eqs. (\ref{wavefunction}), (\ref{BPOrigin}), and (\ref{R}), the
ground-state GP is given formally by
\begin{equation}
\gamma=\frac{-2\pi}{\sqrt{2L^{2}}}\sum_{\mathbf{q}} \mathop{\rm Im} \left(
-A_{\mathbf{q}}e^{2i\mathbf{q}\cdot\mathbf{d}}+A_{\mathbf{q}}^{\ast}\right) ,
\label{BP of R}
\end{equation}
where $\mathbf{d}$\ is a vector from one site to another inside one cell. It
can be seen easily from Fig. \ref{fig1}(a) that the vector is given by $%
\mathbf{d=}\left( \mathbf{n}_{1}+\mathbf{n}_{2}\right) /3$. Inserting this
vector $\mathbf{d}$ into Eq. (\ref{BP of R}), we have%
\begin{align}
\gamma & \!\! =\!\!\frac{-2\pi}{\sqrt{2L^{2}}}\sum_{\mathbf{q}}\mathop{\rm
Im}\left( -A_{\mathbf{q}}e^{\frac{i}{3}\left( 2q_{x}+2q_{y}\right) }+A_{%
\mathbf{q}}^{\ast}\right)  \notag \\
& \!\! =\!\!\frac{-2\pi}{\sqrt{2\!L^{2}}}\!\!\sum_{\mathbf{q}}\!\frac {%
\epsilon_{\mathbf{q}}\!\!\left[ 1\!\!+\!\cos\!\frac{1}{3}\!\left(
2q_{x}\!\!+\!2q_{y}\!\right) \right] \!\!-\!\!\Delta_{\mathbf{q}}\sin \!%
\frac{1}{3}\!\left( 2q_{x}\!\!+\!2q_{y}\right) }{\sqrt{\epsilon _{\mathbf{q}%
}^{2}+\Delta_{\mathbf{q}}^{2}}}  \label{gammafinal}
\end{align}

\begin{figure}[h]
\centering \vspace{0cm} \hspace{0cm} \scalebox{0.5}{%
\includegraphics{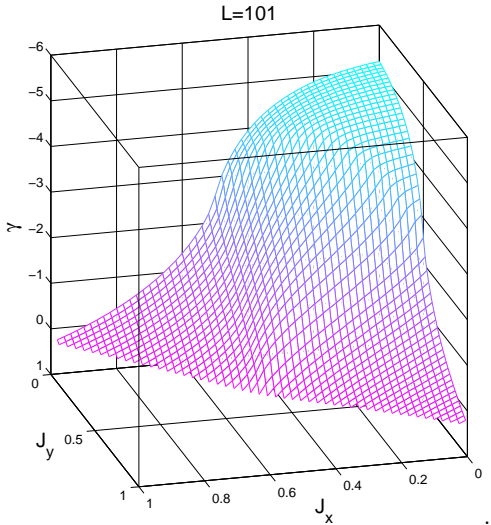}}.
\caption{(Color online) Scaled GP with system size parameter $L=101$.}
\label{fig3}
\end{figure}

\begin{figure*}[t]
\centering \vspace{0cm} \hspace{0cm} \scalebox{1.0}{%
\includegraphics{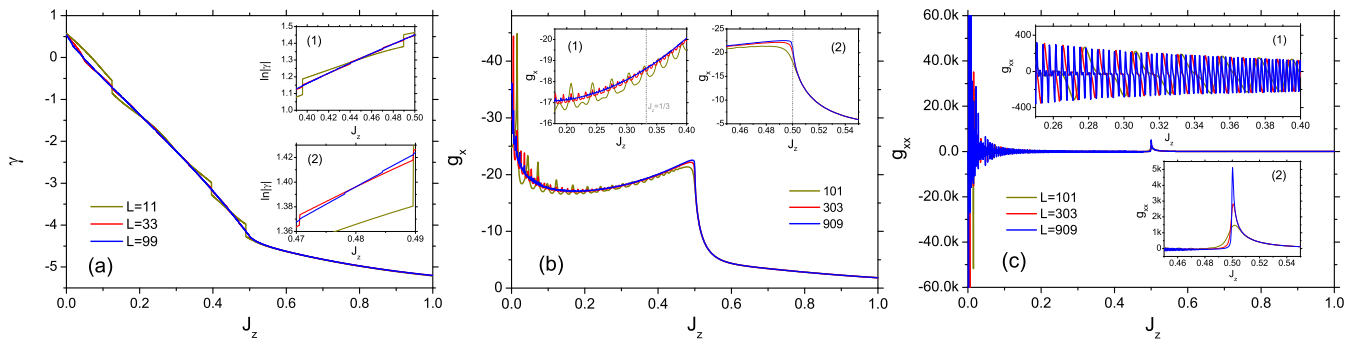}}
\caption{(Color online) (a) $\protect\gamma$ curve along the selected
variation path $J_{x}=J_{y}$ for $L=11,$ $33,$ and $99$. Both insets reveal
the increasing number of saltation in $B$ phase proportional to the system
size parameter $L$. (b) $g_{x}$ and (c) $g_{xx}$ as a function of $J_{z}$
along the variation path $J_{x}=J_{y}$ for system size parameters $L=101,303$
and $909$. The two insets are local enlarged-pictures, which show the
vibration in $B$ phase and the circumstances in the critical point
respectively.}
\label{fig4}
\end{figure*}

Eq. (\ref{gammafinal}) is the main result of this paper. The scaled (or
average) GP $\gamma/L$ as a function of coupling parameters is shown in Fig. %
\ref{fig3}, which distributes symmetrically with respect to the coupling
parameters $J_{x}$ and $J_{y}$. For sake of simplicity, the scaled GP $%
\gamma/L$ is replaced by $\gamma$ in the following discussions. In order to
see the difference between $A$ and $B$ phases, $\gamma$-variation with $%
J_{z} $ along a selected line $J_{x}=J_{y}$ in the phase diagram of Kitaev
\cite{Kitaev} (see Fig. \ref{fig1}(b), the vertical dot-and-dash line) is
plotted in Fig. \ref{fig4}(a). It is shown that $\gamma$ is smooth in $A$
phase ($J_{z}>1/2$) while becomes saltant in $B$ phase ($J_{z}<1/2$) for the
size parameters $L=11$ (dark yellow line)$,$ $33$ (red), and $99$ (blue),
respectively. Moreover, all the data fall onto a single curve in $A$ phase,
while the number of saltation increases with the system-size $L$ in $B$
phase (see insets (1) and (2) of Fig. \ref{fig4}(a) ). To be specific, the
number of saltation for $L=33$ is three times than that for $L=11$ (see Fig. %
\ref{fig4}(a)), and the same situation occurs in turn for $L=99$ and $33$.

It is meaningful to consider the first-order partial derivative $%
g_{_{\beta}}=\partial\gamma/\partial J_{\beta}$ ($\beta=x,y$) of the GP $%
\gamma$. Since the GP $\gamma$ in Eq. (\ref{gammafinal}) is symmetric with
respect to $J_{x}$ and $J_{y}$, we only need to investigate $g_{x}$ (or
equivalently $g_{y}$). The variation of $g_{x}$ with respect to $J_{z}$
along the selected path of $J_{x}=J_{y}$ from $B$ phase to $A$ phase is
shown in Fig. \ref{fig4}(b) for different system-size parameters $L=101$, $%
303$ and $909$. It can be seen from Fig. \ref{fig4}(b) that $g_{x}$
oscillates in $B$ phase with frequency (or number of peaks), which is
proportional to $L$ (see inset (1) of Fig. \ref{fig4}(b)). This rapid
variation of the GP $\gamma$ (in the $B$ phase) has not yet been found, to
our knowledge. However, a very similar behavior of fidelity susceptibility
in the Kitaev model has been reported \cite{S. Yang}. On the other hand, the
value of $g_{x}$ at the critical point $J_{z}=1/2$ increases with the
system-size and sharply decays in $A$ phase (see inset (2) of Fig. \ref{fig4}%
(b) for detail). It is interesting to remark that the saltation of GP $%
\gamma $ in $B$ phase due to the complex structure of ground state with
degeneracy (Fig. \ref{fig2} (b),(c),(d)) is not random rather has
regulation, especially it tends to a regular oscillation above the point, $%
J_{z}=1/3$ , (see inset (1) of Fig. \ref{fig4}(b)). The oscillation
frequency depends linearly on the system size.

To show the non--analyticity of GP at the critical points explicitly the
second-order derivative $g_{xx}$ of the GP $\gamma$ with respect to the
coupling parameters is calculated. Fig. \ref{fig4}(c) shows the variation of
$g_{xx}$ with respect to $J_{z}$ along the variation path of $J_{z}=J_{y}$
for different system-size parameters $L=101$, $303$ and $909$. Inset (1)
reveals the increase of peak-number of $g_{xx}$ in $B$ phase along with the
system-size $L$ similar to $\gamma$ and $g_{x}$ in behavior. The
second-order derivative $g_{xx}$ is divergent at the critical point $%
J_{z}=1/2$ as shown in inset (2) indicating that the TQPT of the Kitaev
honeycomb model is a second-order transition, While the QPT of the XY spin
chain \cite{Zhu,Zhu08} and the Dicke model \cite{Zhu08,Chen} has been shown
to be the first-order transition with the divergent first-order derivative
of the GP. We conclude that the non-analytic GP $\gamma$ can very well
describe the TQPT in terms of the Landau phase-transition theory.

\begin{figure}[h]
\centering \vspace{0cm} \hspace{0cm} \scalebox{0.3}{%
\includegraphics{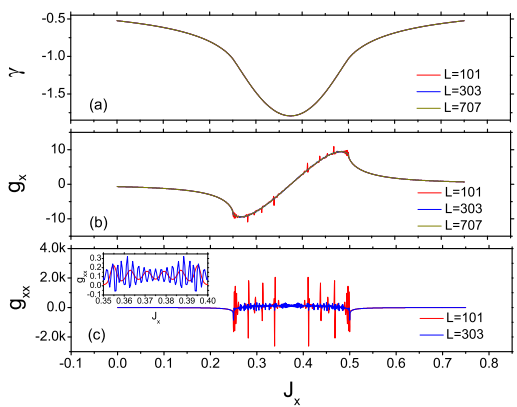}}
\caption{(Color online) (a) $\protect\gamma$ curve, which is smooth in $A$
phase and saltant in $B$ phase, (b) $g_{x}$ curve and (c) $g_{xx}$ as a
function of $J_{x}$ for $J_{x}$ varying along the path $J_{z}=1/4$, with
critical points at $J_{x}=1/4$ and $J_{x}=1/2$.}
\label{fig5}
\end{figure}

Similarly, we can also choose the variation path as $J_{z}=1/4$ (dashed line
in Fig. \ref{fig1}(b)) with two critical points $J_{x}=1/4$ and $J_{x}=1/2$.
Qualitatively similar results are shown in Fig. \ref{fig5} for different
size parameters $L=101$ (red line), $303$ (blue) and $707$ (dark),
respectively, where $\gamma$-plot is a smooth curve in $A$ phase for $%
J_{x}<1/4$ or $J_{x}>1/2$ and becomes saltant in $B$ phase when $%
1/4<J_{x}<1/2$ (Fig. \ref{fig5}(a)). The plots of $g_{x}$ and $g_{xx}$ are
respectively shown in Fig. \ref{fig5}(b) and (c) displaying the clear
vibration in the $B$ phase. $g_{x}$ has a sharp peak at the critical points
and $g_{xx}$ is divergent in the thermodynamic limit showing the
characteristic of second-order phase transition. Inset of Fig. \ref{fig5}(c)
shows the detail of the vibration in the $B$ phase with the peak-number
proportional to the system size.

\section{Finite-size scaling}

\begin{figure}[h]
\centering \vspace{0cm} \hspace{0cm} \scalebox{1.0}{%
\includegraphics{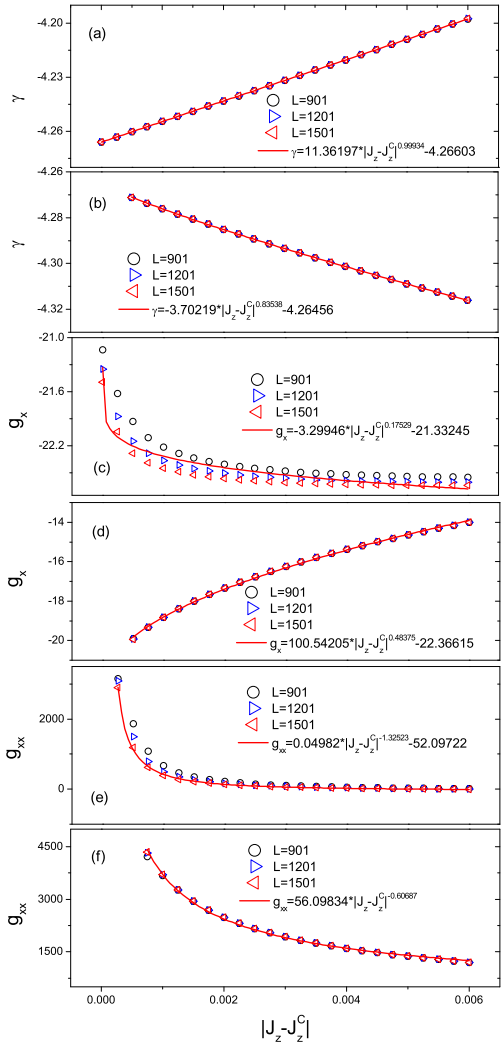}}
\caption{(Color online) Finite-size scaling analysis of the power-law
divergence for (a) GP $\protect\gamma$, (c) $g_{x}$, (e) $g_{xx}$ as a
function of $|J_{z}-J_{z}^{C}|$ in the vicinity of critical point with
system sizes $L=301,901$ and $1501$ on the left-hand side ($J_{z}<J_{z}^{C}$%
) and (b), (d), (f) on the right-hand side ($J_{z}>J_{z}^{C}$) respectively.}
\label{fig6}
\end{figure}

In order to quantify the non-analytic nature of GP through the critical
points, and to further understand the quantum criticality, we investigate
the scaling behavior of the GP and its first-order and second-order
derivatives by the finite size scaling analysis \cite{Barber}. Since the GP
of a many-body system is an extensive quantity, it usually depends on the
system size in the non-critical region. In the vicinity of critical point ($%
J_{z}^{C}=1/2$ here), the GP $\gamma$ follows a power law\cite{Lin}

\begin{equation}
\gamma \propto {\left\vert J_{z}-J_{z}^{C}\right\vert }^{-\alpha _{\gamma }}
\label{alpha}
\end{equation}%
with $\alpha $ being corresponding exponent. ${\left\vert
J_{z}-J_{z}^{C}\right\vert }$-dependence of $\gamma $ in the left-hand side
of critical point ( $J_{z}<J_{z}^{C}$ ) and right-hand side ( $%
J_{z}>J_{z}^{C})$ is respectively plotted in Fig. \ref{fig6}(a) and (b) for
different system-size parameters $L=101$, $303$ and $909$. The corresponding
exponents $\alpha _{\gamma }^{-}=-0.99934\pm 0.00033$ (left-hand side) and $%
\alpha _{\gamma }^{+}=-0.83538\pm 0.00008$ (right-hand side) are obtained
from Fig. \ref{fig6}(a) and (b). Similarly, $g_{x}$ and $g_{xx}$ as a
function of $|J_{z}-J_{z}^{C}|$ are plotted in Fig. \ref{fig6}(c), (d) and
(e), (f) for the gapless and gapped phases respectively, from which the
critical exponents $\alpha _{g_{x}}^{-}=-0.17529\pm 0.02056$, $\alpha
_{g_{xx}}^{-}=1.32523\pm 0.01719$ (left) for B phase and $\alpha
_{g_{x}}^{+}=-0.48375\pm 0.00033$, $\alpha _{g_{xx}}^{+}=0.60687\pm 0.00526$
(right) for A phase are found. The fact of negative exponents $\alpha
_{\gamma }^{\pm }$, $\alpha _{g_{x}}^{\pm }$ and positive $\alpha
_{g_{xx}}^{\pm }$ indicates that $\gamma $ and $g_{x}$ are finite while $%
g_{xx}$ is divergent at the critical point in the thermodynamic limit. Thus
the TQPT is a second-order phase transition characterized by the GP $\gamma $%
.

\begin{figure}[h]
\centering \vspace{0cm} \hspace{0cm} \scalebox{0.35}{%
\includegraphics{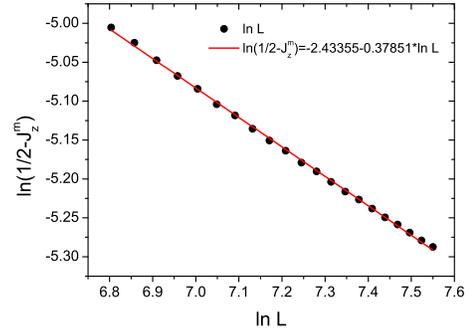}}
\caption{(Color online) $L$-dependence of $J_{z}^{C}-J_{z}^{m}$ in
logarithmic coordinate for $L=901,951,...,1901$.}
\label{fig7}
\end{figure}

The position of maximum value of $\gamma $ denoted by $J_{z}^{m}$ may not be
located exactly at the critical point $J_{z}^{C}=1/2$, but tends to it in
the thermodynamic limit $L\rightarrow \infty $, which is regarded as the
pseudocritical point \cite{Barber}. The $J_{z}^{C}-J_{z}^{m}$ versus
different system-size $L=901,951,...,1901$ in a logarithmic coordinate is
plotted in Fig. \ref{fig7}, which is a straight line of slope $-0.37851\pm
0.00226$. It means that the $J_{z}^{m}$ tends toward to the critical point $%
J_{z}^{C}$ following the power-law decay
\begin{equation}
J_{z}^{C}-J_{z}^{m}\propto L^{-0.37851}.
\end{equation}

\begin{figure*}[t]
\centering \vspace{0cm} \hspace{0cm} \scalebox{1.0}{%
\includegraphics{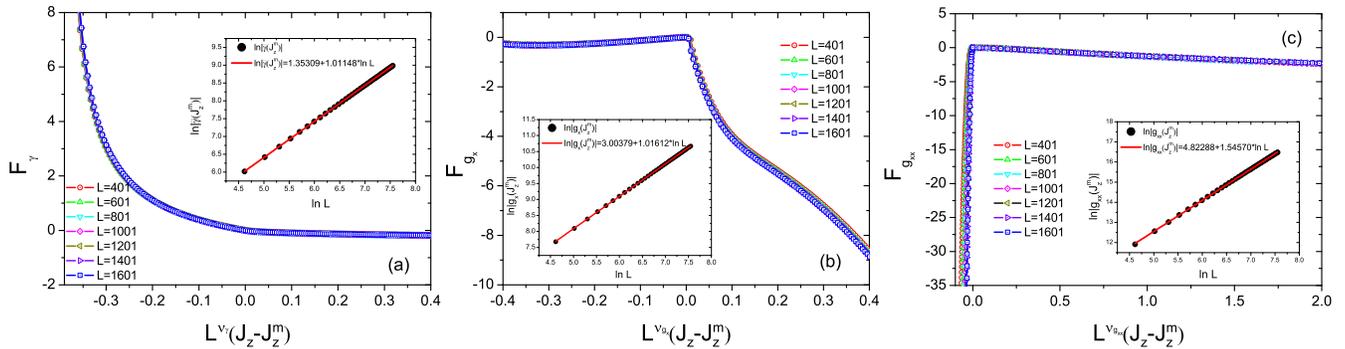}}
\caption{(Color online) (a) $F_{\protect\gamma}$, (b) $F_{g_x}$, and (c) $%
F_{g_{xx}}$ as a function of $L^{\protect\nu}\left( J_{z}-J_{z}^{m}\right)$
for $L=401,601,...,1601$. All the data fall on to a single curve
respectively. And the three insets in turn shows the variation of $\protect%
\gamma(J_{z}^{m})$, $g_{x}(J_{z}^{m})$, and $g_{xx}(J_{z}^{m})$ with respect
to the system size-parameters $L=101,151,...,1901$.}
\label{fig8}
\end{figure*}

On the other hand, the maximum value of $\gamma $ at $J_{z}=J_{z}^{m}$ for a
finite-size system behaves as
\begin{equation}
\gamma ({J_{z}^{m}})\propto {L}^{\mu _{\gamma }},
\end{equation}%
which is shown in the inset of Fig. \ref{fig8}(a) with a straight line in
logarithmic coordinate. The corresponding size-exponent is given by $\mu
_{\gamma }=0.01148\pm 0.00054$.

Since the GP $\gamma$ around its maximum position $J_{z}^{m}$ can be written
as a simple function of $J_{z}^{m}-J_{z}$, it is possible to make all the
value-data defined by a universal scaling function $F_{\gamma}=\left( {\gamma%
}({J_{z}^{m}})-\gamma\right) /\gamma$ versus $L^{\nu_{\gamma}}\left(
J_{z}-J_{z}^{m}\right) $, namely,
\begin{equation}
F_{\gamma}=f\left[ L^{\nu_{\gamma}}\left( J_{z}^{m}-J_{z}\right) \right] ,
\end{equation}
where $\nu_{\gamma}$ is a critical exponent that governs the divergence of
the correlation length. The values of $F_{\gamma}$ for different system-size
parameters $L$ fall onto a single curve as shown in Fig. \ref{fig8}(a), from
which we can extract the critical exponent $\nu_{\gamma}=-0.015$ numerically.

In fact, according to the scaling ansatz of a finite system \cite{Barber,Lin}%
, the critical exponent $\nu $ can be determined by the relation $\nu =\mu
/\alpha $. In terms of this relation, the critical exponents in $B$ and $A$
phases are found as ${\nu }_{\gamma }^{-}=-0.01149$ and ${\nu }_{\gamma
}^{+}=-0.01374$, which are consistent with the numerical result $\nu
_{\gamma }$ extracted from Fig. \ref{fig8}(a). Inset of Fig. \ref{fig8}(b)
is a plot of the maximum value of $g_{x}({J_{z}^{m}})$ as a function of $L$
\begin{equation}
g_{x}({J_{z}^{m}})\propto {L}^{\mu _{g_{x}}}
\end{equation}
in logarithmic coordinate, from which the size exponent $\mu
_{g_{x}}=0.01612\pm 0.00094$ is found. The universal scaling function
\begin{equation}
F_{g_{x}}=f\left[ L^{\nu _{g_{x}}}(J_{z}^{m}-J_{z})\right]
\end{equation}
is shown in Fig. \ref{fig8}(b). We have the numerical value $\nu
_{g_{x}}=-0.040$ and the results determined by the relation $\nu =\mu
/\alpha $ that $\nu _{g_{x}}^{-}=-0.09196$, $\nu _{g_{x}}^{+}=-0.03332$. The
deviation between $\nu _{g}$ and $\nu _{g_{xx}}^{-}$ may be due to the rapid
oscillation in the gapless $B$ phase. Similarly, the critical exponents of $%
g_{xx}$ can be obtained from Fig. \ref{fig8}(c) as $\mu _{g_{xx}}=0.54570\pm
0.00237$, $\nu _{g_{xx}}^{-}=0.410$ and $\nu _{g_{xx}}^{+}=0.908$, and the
results determined by $\nu =\mu /\alpha $ are $\nu _{g_{xx}}^{-}=0.41178$
and $\nu _{g_{xx}}^{+}=0.89920$ respectively.

\section{Summary and discussion}

We demonstrate that the ground-state GP generated by the correlated rotation
of two linked-spins in a unit-cell indeed can be used to characterize the
TQPT for the Kitaev honeycomb model. The non-analytic GP with a divergent
second-order derivative at the critical points shows that the TQPT is a
second-order phase-transition different from the $XY$ spin-chain \cite%
{Zhu,Zhu08}, in which the first-order derivative of GP is divergent, and the
LMG model\cite{Zhu,Zhu08}, in which the GP itself is shown to be divergent.
Moreover it is found that the GP is zigzagging with oscillating derivatives
in the gapless $B$ phase, but is a smooth function in the gapped $A$ phase.
The scaling behavior of the non-analytic GP in the vicinity of critical
point is shown to exhibit the universality with negative exponents of both $%
\gamma$ and $g_{x}$ while a positive exponent of $g_{xx}$ indicating the
characteristic of second-order phase transition.

\section*{Acknowledgments}

This work is supported by the NNSF of China under Grant Nos. 11075099 and
11074154, ZJNSF under Grant No. Y6090001, and the 973 Program under Grant
No. 2006CB921603.

\end{document}